\documentclass[prl,twocolumn,superscriptaddress]{revtex4}
\usepackage{graphicx}

\newcommand{ \ybco }{\mbox{YBa$_2$Cu$_3$O$_{6+x}$}}
\newcommand{ \ybcoss }{\mbox{YBa$_2$Cu$_3$O$_{6.6}$}}

\newcommand{ \qaf }{\mbox{\textbf{Q}$_{\text AF}$}}

\newcommand{ \astar }{\mbox{$a^*$}}
\newcommand{ \bstar }{\mbox{$b^*$}}
\newcommand{ \cstar }{\mbox{$c^*$}}

\newcommand{ \Eres }{\mbox{$E_{\text res}$}}

\newcommand{ \IC }{\mbox{$\delta$}}

\begin{document}

\title{Spin dynamics in the pseudogap state of a high-temperature superconductor}

\author{V. Hinkov}
\affiliation{Max-Planck-Institut f{\"u}r Festk{\"o}rperforschung, Heisenbergstra{\ss}e 1, 70569 Stuttgart, Germany}

\author{P. Bourges}
\affiliation{Laboratoire L{\'e}on Brillouin, CEA-CNRS, CE-Saclay, 91191 Gif-sur-Yvette, France}

\author{S. Pailhes}
\affiliation{Laboratoire L{\'e}on Brillouin, CEA-CNRS, CE-Saclay, 91191 Gif-sur-Yvette, France}

\author{Y. Sidis}
\affiliation{Laboratoire L{\'e}on Brillouin, CEA-CNRS, CE-Saclay, 91191 Gif-sur-Yvette, France}

\author{A. Ivanov}
\affiliation{Institut Laue-Langevin, 6 Rue Jules Horowitz, 38042 Grenoble cedex 9, France}

\author{C. D. Frost}
\affiliation{ISIS Facility, Rutherford Appleton Laboratory, Chilton, Didcot, Oxon, UK}

\author{T. G. Perring}
\affiliation{ISIS Facility, Rutherford Appleton Laboratory, Chilton, Didcot, Oxon, UK}

\author{C. T. Lin}
\affiliation{Max-Planck-Institut f{\"u}r Festk{\"o}rperforschung, Heisenbergstra{\ss}e 1, 70569 Stuttgart, Germany}

\author{D. P. Chen}
\affiliation{Max-Planck-Institut f{\"u}r Festk{\"o}rperforschung, Heisenbergstra{\ss}e 1, 70569 Stuttgart, Germany}

\author{B. Keimer}
\affiliation{Max-Planck-Institut f{\"u}r Festk{\"o}rperforschung, Heisenbergstra{\ss}e 1, 70569 Stuttgart, Germany}

\begin{abstract} The pseudogap is one of the most pervasive
phenomena of high temperature superconductors \cite{Tim99}. It is
attributed either to incoherent Cooper pairing setting in above the
superconducting transition temperature $T_c$, or to a hidden order
parameter competing with superconductivity. Here we use inelastic
neutron scattering from underdoped \ybcoss\ to show that the
dispersion relations of spin excitations in the superconducting and
pseudogap states are qualitatively different. Specifically, the
extensively studied ``hour glass" shape of the magnetic dispersions
in the superconducting state \cite{Hay04,Pai04,Rez04} is no longer
discernible in the pseudogap state and we observe an unusual
``vertical" dispersion with pronounced in-plane anisotropy. The
differences between superconducting and pseudogap states are thus
more profound than generally believed, suggesting a competition
between these two states. Whereas the high-energy excitations are
common to both states and obey the symmetry of the copper oxide
square lattice, the low-energy excitations in the pseudogap state may
be indicative of collective fluctuations towards a state with broken
orientational symmetry predicted in theoretical work
\cite{Kiv98,Hal00,Yam00,Kee03}.

Original reference: Nature Physics 3, 780 (2007).
\end{abstract}

\maketitle


The pseudogap (PG) manifests itself in various experimental probes as
a depletion of spectral weight upon cooling below a doping-dependent
characteristic temperature $T^\ast$. The origin of the pseudogap
remains the focus of significant debate. For instance, recent
experiments have provided evidence of vortex-like excitations below
$T^\ast$ akin to those in the superconducting (SC) state
\cite{Ong03}. Other experiments \cite{Fau06} suggest the presence of
a novel magnetic order breaking time-reversal symmetry, again setting
in around $T^\ast$.

Inelastic neutron scattering directly probes the microscopic magnetic
dynamics and can thus serve as particularly incisive tests of
microscopic models of the cuprates. Recent work on non-superconducting
$\rm La_{2-x} (Sr,Ba)_{x} CuO_4$ and in the SC state of \ybco\ has
uncovered tantalizing evidence of a ``universal" spin excitation
spectrum independent of material-specific details
\cite{Hay04,Tra04,Chr04,Pai04,Rez04}. The dispersion surface
comprises upward- and downward-dispersing branches merging at the
wave vector \qaf, which characterizes antiferromagnetic order in the
undoped parent compounds. The spectrum thus resembles an ``hour
glass'' in energy-momentum space. Investigations of the SC state of
\emph{twin-free} \ybco\ (i.e., single crystals with a unique
orientation of the two in-plane axes $a$ and $b$ throughout the
entire volume) have revealed a two-dimensional geometry of the lower
branch, with a modest in-plane anisotropy of the spectral weight that
increases with decreasing excitation energy \cite{Hin04}. The ``hour
glass'' spectrum and its in-plane anisotropy have stimulated
theoretical work in the framework of both itinerant and local-moment
pictures of the electron system
\cite{Sch06,Sei05,Uhr04,Voj06,Yao06,Yam06}.

Previous measurements of twinned \ybco-samples appeared to indicate
that the spin excitation spectrum in the PG state is simply a
broadened version of the spectrum in the SC state
\cite{Dai99,Fon00,Sto04}. However, the detailed investigation of the
geometry and dispersion of spin excitations in the PG state reported
here reveals that both spectra are distinctly different.

\begin{figure}[]
\includegraphics[width=8.5cm]{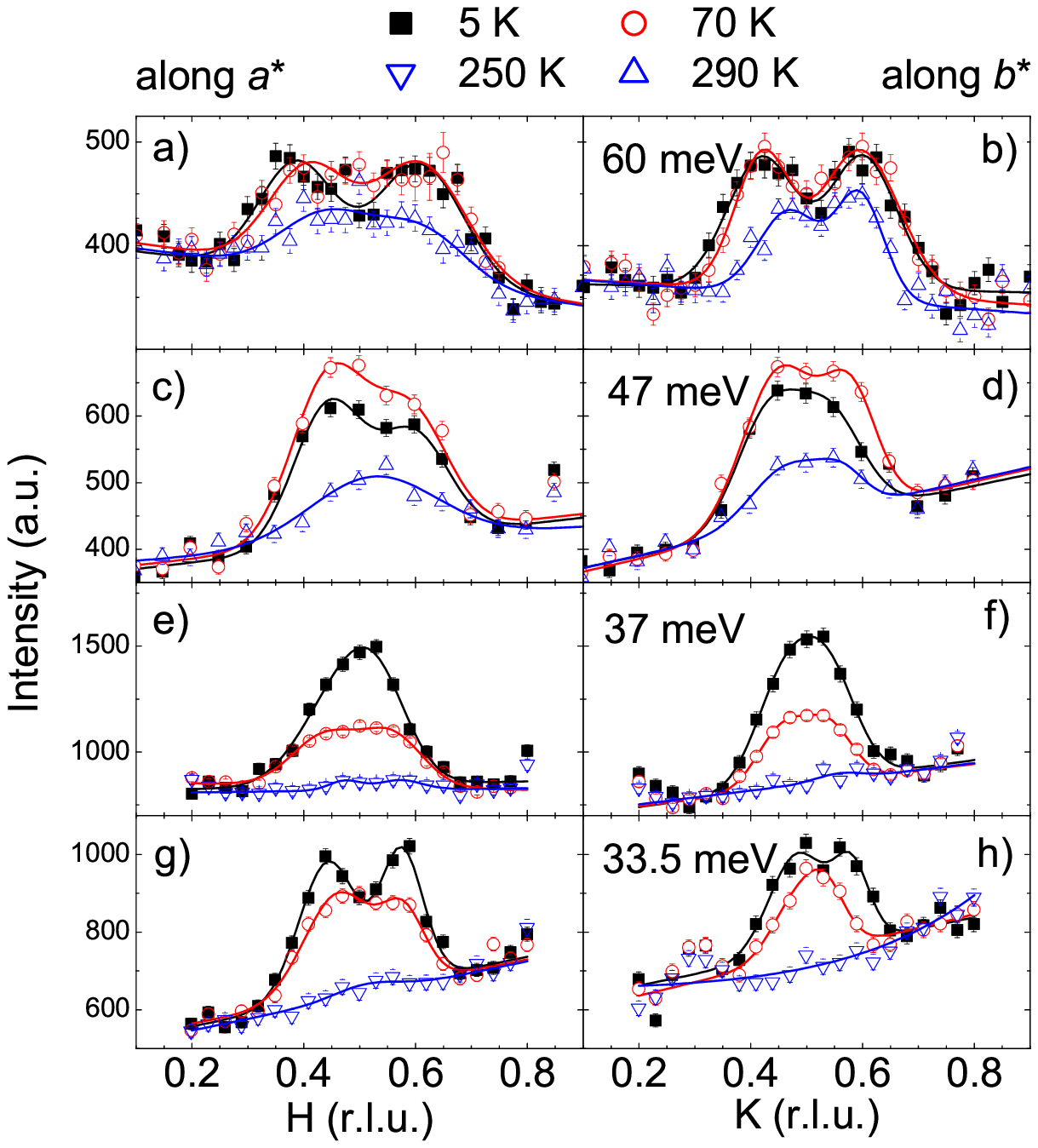}
\caption{Energy evolution of the in-plane magnetic excitations around
\qaf\ for different temperatures. In \textbf{a,b}, the energy transfer was fixed to \mbox{60 meV}, in \textbf{c,d} to \mbox{47 meV}, in \textbf{e,f}
to \mbox{37.5 meV} and in \textbf{g,h} to \mbox{33.5 meV}. Panels \textbf{a,c,e,g} show scans along the $a$-axis and panels \textbf{b,d,f,h} scans along the $b$-axis. The lines are the results of fits to Gaussian profiles. We show the raw triple-axis data; the only data processing applied is a subtraction
of a \emph{constant} at 250~K and 290~K in order to account for the increased background from multi-phonon scattering. Corrections for the Bose factor are
small and were not applied to the data. The final wave vector was fixed to $2.66 \mbox{ \AA}^{-1}$ below 38~meV and to $4.5 \mbox{ \AA}^{-1}$ above. Error-bars indicate the statistical error.
}
\label{figRaw3ax}
\end{figure}

The measurements were performed on an array of twin-free single
crystals of underdoped \ybcoss\ with superconducting $\rm T_c = 61$~K
(see Methods). Fig. 1 shows representative raw data in the SC state
at 5~K, in the PG state just above $T_c$, and at room temperature.
The incommensurability \IC\ and the spectral weight of the
constant-energy cuts are generally different in the two in-plane
directions in reciprocal space, $H$ and $K$, and show a complex
dependence upon energy and temperature (see Methods for definitions).
A synopsis of the entire data set is shown in Fig. 2 at two
temperatures above and below $T_c$.

\begin{figure}[b]
\includegraphics[width=8.5cm]{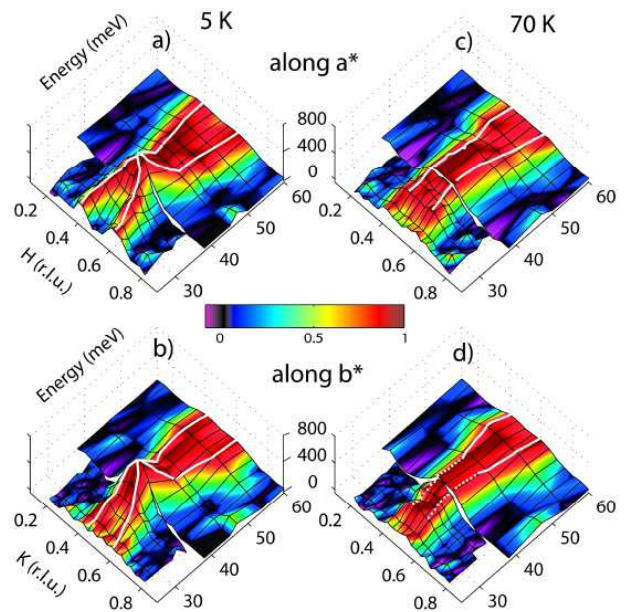}
\caption{ Colour representation of the magnetic intensity, obtained from triple-axis scans. Panels \textbf{a,b} show the SC regime and \textbf{c,d} the regime just above $T_c$. The upper and lower rows show scans along the $a$-axis ($H$, -1.5, -1.7) and $b$-axis (1.5, $K$, 1.7), respectively. In order to obtain a meaningful colour representation, the intensity at 250 K was subtracted for $E <$ 38~meV and the data was corrected for a \textbf{Q}-linear background
at all energies. At each individual energy, the colour scale was normalized
to the peak intensity of the scan, allowing a better comparison of the Q-extent
at different energies. The final wave-vector was fixed to $2.66 \mbox{ \AA}^{-1}$ below 38~meV and to $4.5 \mbox{ \AA}^{-1}$ above. Scans taken at the overlapping energy 38~meV were used to bring both energy ranges to the same scale. Crossings of black lines represent measured data points. White lines connect the fitted peak positions of the constant-energy cuts. Dotted lines represent upper bounds on the incommensurability.
}
\label{figColor3ax}
\end{figure}

The central result of this paper is the change of the topology of the
dispersion surface from the SC to the PG state. First, we focus on
the spectrum in the SC state (Fig. 1, 5~K). Starting from low
excitation energies, \IC\ first decreases with increasing energy, so
that the incommensurate peaks merge at \qaf\ at an energy of
$\Eres=37.5$ meV. For $E > \Eres$, \IC\ increases again, so that the
spectrum forms the ``hour-glass" dispersion (Fig. 2\textbf{a,b})
already familiar from previous work \cite{Hay04,Tra04,Bou00,Rez04}.
In the PG state, however, the singularity at \Eres\ is no longer
discernible, and \IC\ is only weakly energy-dependent over a wide
energy range including \Eres\ (Fig. 2\textbf{c,d} and Supplementary
Information). This constitutes a qualitative difference of the spectra in
the SC and PG states.

The spin excitations in both states also differ markedly with respect
to their \emph{a-b}-anisotropy. In the SC state, constant-energy cuts
of the magnetic spectral weight at energies below $\Eres$ form
ellipses with aspect ratios of about 0.8 (Figs. 1\textbf{g,h} and
3\textbf{a}). In the PG state, the $Q$-extent of the signal decreases
significantly along \bstar\ such that incommensurate peaks can hardly
be resolved, Fig. 1\textbf{h}. In contrast, the flat-top profiles
along \astar\ are well described by two broad peaks displaced from
\qaf, Fig. 1\textbf{g}, and we can set an upper bound of 0.6 on the
ratio of \IC\ along \bstar\ and \astar. The intensity distribution of
the high-energy excitations, on the other hand, hardly changes
between the SC and PG states (Figs. 1\textbf{a,b} and 3\textbf{b}).
It is also much more isotropic than the low-energy profiles.
Specifically, the ratio of \IC\ along \astar\ and \bstar\ is
$0.95\pm0.07$ as obtained from an analysis of the time-of-flight and
triple axis data. Moreover, there is no amplitude suppression along
\bstar\ and the anisotropy in the peak widths is also less pronounced
than at low energies ($0.16\pm0.02$ and $0.13\pm0.02$ r.l.u. along
\astar\ and \bstar, respectively).

\begin{figure}[]
\includegraphics[width=8.5cm]{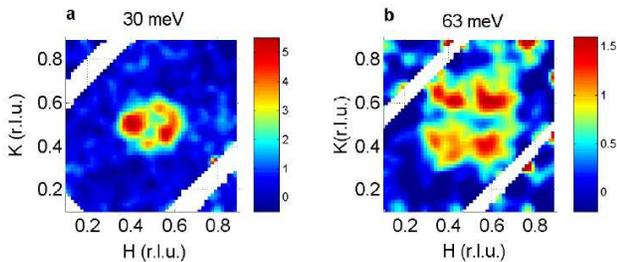}
\caption{Colour representation of the in-plane magnetic intensity collected at the TOF spectrometer at 5~K. The incident energy was fixed to $E_i=120$~meV and the counting time was 48 hours. In \textbf{a} intensity was integrated over the energy range $E=(30 \pm 3.5)$~meV and in \textbf{b} over $E=(63\pm
5)$~meV. The data represent acoustic excitations and were collected close to the maximum $L$-component at $L=2.1$ and 4.7, respectively (see Methods). A background quadratic in \textbf{Q} was subtracted and data are shown in arbitrary units.
}
\label{figISIS}
\end{figure}

Upon cooling below $T_c$, the spectral rearrangement associated with
the formation of the downward-dispersing branch of the ``hour-glass"
results in a sharp upturn of the intensity at points along this
branch (Fig. 4\textbf{a,b}), while at \qaf\ and 30~meV there is only
a broad maximum at $T_c$ (Fig. 4\textbf{c}). This is a further
manifestation of the qualitative difference between the SC and PG
state spectra. In the PG state, the spectral weight at energies at
and below \Eres\ declines uniformly with increasing temperature at
all \textbf{Q}-values and vanishes
around $T^\ast \approx$ 200~K. $T^\ast$ is comparable to the
temperature below which the PG manifests itself in other experimental
probes \cite{Tim99,Ong03,Fau06}. Corresponding constant-energy cuts
show that for $T > T^*$, the low-energy spectral weight is severely
depleted over the entire Brillouin zone below an
energy $E^* \sim 40$ meV, Fig. 1\textbf{e-h} (see also Supplementary
Notes and Ref. \cite{Dai99}). In contrast, at energies above $E^*$
the spectral weight is only moderately reduced upon heating above
$T^*$ (Fig. 1\textbf{a,b}). We can thus distinguish a third
temperature regime above $T^*$, with a magnetic excitation spectrum
differing distinctly from the spectra in the SC state and just above
$T_c$.

Two aspects of our data in the PG state merit further comment. First,
we observe the low-energy magnetic spectral weight to increase upon
cooling below $T^\ast$, in contrast to the loss of spectral weight
indicated by nuclear magnetic resonance (NMR) data \cite{Tim99}.
However, one needs to keep in mind that the energy scales probed by
NMR and neutron scattering differ by three order of magnitude. The
behaviour we observe can be viewed as a pile-up of spectral weight
above a spin-pseudogap of $\approx 15$ meV, which is much larger than
the NMR energy scale but consistent with prior neutron scattering
work on underdoped \ybco\ \cite{Fon00}. Second, the magnitude of the
spin-PG implied by our data is much lower than the charge-PG observed
in infrared spectroscopy \cite{Tim99}. The divergence of gap features
in spin and charge channels is not unexpected in a correlated
metallic state close to the Mott insulator (where the spin gap
vanishes, and the charge gap is $\sim 2$ eV).

\begin{figure}[b]
\includegraphics[width=8.5cm]{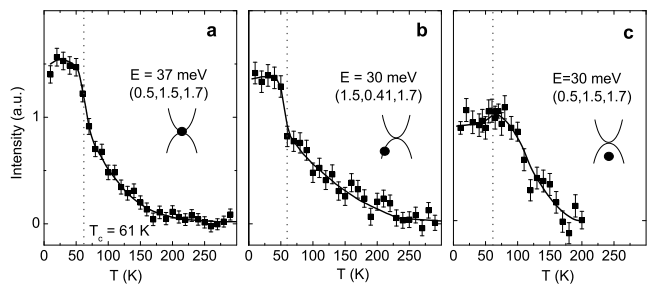}
\caption{Temperature dependence of the peak intensity at three different positions in energy-momentum space sketched in reference to the ``hour glass'' dispersion in the SC state. In \textbf{a}, the energy transfer was fixed to $E=37\text{ meV}$ and the wave-vector transfer to $\mathbf{Q}=(0.5,1.5,1.7)$.
In \textbf{b}, $E=30\text{ meV}$ and $\mathbf{Q}=(1.5,0.41,1.7)$ and in \textbf{c},
$E=30\text{ meV}$ and $\mathbf{Q}=(0.5,1.5,1.7)$. A weighted average of the background measured at two representative points was subtracted. The units
in the three panels were scaled to approximately the same intensity at $T_c$.
Error-bars indicate the statistical error.
}
\label{figTempDep}
\end{figure}

We now discuss the implications of our observations for the
microscopic description of the PG state. Many features of the
magnetic spectrum in the SC state are well described by models that
regard the downward-dispersing branch as an excitonic mode in the
spin-triplet channel below the SC energy gap
\cite{Esc05,Sch06,Pai06}. If the downward dispersion is a
manifestation of the d-wave symmetry of the gap, it is expected to
disappear in the normal state, as observed. However, our observation
of coherent, highly anisotropic spin excitations for $T_c < T < T^*$
requires a fresh theoretical approach. Specifically, the striking
difference between the magnetic dynamics in the SC and PG states
implies that these excitations cannot simply be regarded as
incoherent precursors of the resonant excitations below $T_c$.
Rather, they appear to be a signature of a many-body state that
competes with superconductivity.

The strong in-plane anisotropy of the spin dynamics in the PG state
offers intriguing clues to the nature of this state. The fourfold
orientational symmetry of the CuO$_2$ planes is broken in \ybco\ by
the CuO chains running along the $b$-direction. However,
weak-coupling calculations assuming a Fermi liquid state above $T_c$
in conjunction with recent ARPES-measurements \cite{Zab06} indicate
that the hybridization between electronic states on CuO chains and
CuO$_2$ layers is not large enough to explain the in-plane anisotropy
at the low-energy end of the magnetic spectrum in the SC state
\cite{Sch06}. Based on these model calculations \cite{Sch06}, it is
quite unlikely that band structure effects alone can explain the much
larger in-plane anisotropy and the strong temperature dependence we
have observed above $T_c$. The low-energy spin dynamics in the PG
state thus appears to reflect a susceptibility of the two-dimensional
electron system to a weak in-plane symmetry-breaking field (induced
by the CuO chains) that is markedly larger than that of a weakly
renormalized Fermi liquid. Two possibilities have been discussed
recently: (i) a ``striped" state with quasi-one-dimensional order of
spins and charges that breaks rotational and translational symmetry
\cite{Tra95, Kiv03}; and (ii) a ``nematic" or ``Pomeranchuk" state
that breaks the rotational symmetry only
\cite{Kiv98,Hal00,Yam00,Kee03}. Experimental evidence for rotational
symmetry breaking has also been gleaned from transport experiments
\cite{And02}.

A static striped state can be ruled out for \ybcoss\ based on the
two-dimensional geometry of the high-energy spin excitations (Fig.
3), and on the absence of Bragg reflections indicative of static spin
and charge order. However,
predictions for the spin dynamics of fluctuating stripe \cite{Voj06}
and Pomeranchuk states \cite{Yam06} bear some resemblance to our
data. Specifically, fluctuating stripe segments with correlations
lengths of the order of a few lattice spacings can result in an
approximately ``vertical", highly anisotropic dispersion surface at
low energies, while the high-energy excitations are isotropic, as
observed. However, calculations that explicitly consider the
influence of superconductivity on the spin excitations of striped
states \cite{And05} show only a rather modest effect (namely, the
opening of a spin gap), in disagreement with our observations.
Calculations for itinerant electron systems close to a Pomeranchuk
instability \cite{Yam06} are also in good overall agreement with our
data; however in the SC state, the low-energy anisotropy of the
spectral weight is significantly weaker than observed. Despite these
promising first steps, an explanation of the unusual spin dynamics we
have observed
therefore remains an interesting challenge for theory. Its
relationship to other fingerprints of the PG state, such as the
presence of vortex-like excitations \cite{Ong03} and unconventional
magnetic order \cite{Fau06}, is also not understood at present.

The full in-plane momentum resolution we have now achieved also opens
a new window on the universality of the spin excitations in the
cuprates. The high-energy spin excitations with four peaks along the
diagonals of the CuO$_2$ plaquettes are indeed common to all layered
cuprate compounds thus far investigated. These excitations do not
show marked changes at $T_c$, and their intensity distribution
exhibits at most a weak in-plane anisotropy. Our data show that the
geometry and spectral weight distribution at lower energies depends
more strongly on materials and temperature. In particular, we have
shown that the spectra in the normal and superconducting states are
qualitatively different.
These differences may reflect the competition between
high-temperature superconductivity and other ordering phenomena such
as ``nematic" or ``Pomeranchuk" states.

Correspondence and requests for materials should be addressed to B.K.
(b.keimer@fkf.mpg.de).

\textbf{Acknowledgements}

We thank C. Bernhard, G. Khaliullin, J. Lorenzana, D. Manske, W.
Metzner, G. Seibold, M. Vojta, and H. Yamase for stimulating
discussions and M. Raichle and M. Br\"{o}ll for support with the
sample preparation. This work was supported in part by the Deutsche
Forschungsgemeinschaft in the consortium FOR538.

\textbf{Competing financial interests}

The authors declare that they have no competing financial interests.

\section*{Methods}

The experiments were performed on an array of 180 individually
detwinned \ybcoss\ crystal with superconducting transition
temperatures (midpoint) of $T_c \approx 61$~K and width $\Delta T_c
\approx 2$~K, determined for each crystal by magnetometry
\cite{Hin04}. They were co-aligned on three Al-plates with a
mosaicity of $<1.2 ^\circ$. The volume of the entire array was $\sim
450$~mm$^3$, and the twin domain population ratio was 94:6.

Triple-axis measurements (Figs. 1, 2 and 3) were performed at the IN8
spectrometer at the Institut Laue Langevin (Grenoble, France) and the
2T spectrometer at the Laboratoire L\'{e}on Brillouin (Saclay,
France). At both instruments, horizontally and vertically focusing
crystals of pyrolytic graphite, set for the (002) reflection, were
used to monochromate and analyze the neutron beam. Scans along
\astar\ and \bstar\ were carried out under identical instrumental
resolution conditions by working in two different Brillouin zones,
($H$, -1.5, -1.7) and (1.5, $K$, 1.7). No collimators were used in
order to maximise the neutron flux, and graphite filters extinguished
higher order contamination of the neutron beam. Extensive resolution
calculations confirm that the effects discussed in the main text are
no resolution effects. Specifically, the spectra at 70K exhibit only
structures that are significantly broader than the resolution
function, so that the observed spectrum is almost undistorted by
resolution effects.

Time-of-flight measurements (Fig. 3) were performed at the
MAPS-spectrometer of the ISIS spallation source, Rutherford Appleton
Laboratory (Chilton, UK). The source proton current was $170\; \mu$A
and the Fermi chopper rotation frequency was set to 250~Hz.

The wave vector is quoted in units of the reciprocal lattice vectors
\astar, \bstar\ and \cstar\, where $a=2 \pi / \astar =3.82 \mbox{
\AA}$, $b=3.87 \mbox{ \AA}$ and $c=11.7 \mbox{ \AA}$. We choose its
out-of-plane component $L_0 = 1.7 \times (2n+1)$, $n$ integer, to
probe magnetic excitations that are odd under the exchange of two
layers within a bilayer unit. As even excitations show a gap of
$\approx 55$ meV and are much less $T$-dependent, they are presented
elsewhere \cite{Pai06}. The incommensurability \IC\ is defined as the
deviation of the peak position from $\qaf=(0.5, 0.5)$. Even in the
case, where two incommensurate peaks are close together or broad
(``flat-top'' structure), \IC\ can be obtained by fitting Gaussians
to the data.

\bibliographystyle{naturemag}

\begin{thebibliography}{10}
\expandafter\ifx\csname url\endcsname\relax
  \def\url#1{\texttt{#1}}\fi
\expandafter\ifx\csname urlprefix\endcsname\relax\def\urlprefix{URL }\fi
\providecommand{\bibinfo}[2]{#2}
\providecommand{\eprint}[2][]{\url{#2}}

\bibitem{Tim99}
\bibinfo{author}{Timusk, T.} \& \bibinfo{author}{Statt, B.~W.}
\newblock \bibinfo{title}{The pseudogap in high-temperature superconductors: an
  experimental survey}.
\newblock \emph{\bibinfo{journal}{Rep. Prog. Phys.}}
  \textbf{\bibinfo{volume}{62}}, \bibinfo{pages}{61--122}
  (\bibinfo{year}{1999}).

\bibitem{Hay04}
\bibinfo{author}{Hayden, S.~M.}, \bibinfo{author}{Mook, H.~A.},
  \bibinfo{author}{Dai, P.}, \bibinfo{author}{Perring, T.~G.} \&
  \bibinfo{author}{Dogan, F.}
\newblock \bibinfo{title}{The structure of the high-energy spin excitations in
  a high-transition-temperature superconductor}.
\newblock \emph{\bibinfo{journal}{Nature}} \textbf{\bibinfo{volume}{429}},
  \bibinfo{pages}{531--534} (\bibinfo{year}{2004}).

\bibitem{Pai04}
\bibinfo{author}{Pailh\`{e}s, S.} \emph{et~al.}
\newblock \bibinfo{title}{Resonant magnetic excitations at high energy in
  superconducting {YBa$_2$Cu$_3$O$_{6.85}$}}.
\newblock \emph{\bibinfo{journal}{Phys. Rev. Lett.}}
  \textbf{\bibinfo{volume}{93}}, \bibinfo{pages}{167001}
  (\bibinfo{year}{2004}).

\bibitem{Rez04}
\bibinfo{author}{Reznik, D.} \emph{et~al.}
\newblock \bibinfo{title}{Dispersion of magnetic excitations in optimally doped
  superconducting {YBa$_2$Cu$_3$O$_{6.95}$}}.
\newblock \emph{\bibinfo{journal}{Phys. Rev. Lett.}}
  \textbf{\bibinfo{volume}{93}}, \bibinfo{pages}{207003}
  (\bibinfo{year}{2004}).

\bibitem{Kiv98}
\bibinfo{author}{Kivelson, S.~A.}, \bibinfo{author}{Fradkin, E.} \&
  \bibinfo{author}{Emery, V.~J.}
\newblock \bibinfo{title}{Electronic liquid-crystal phases of a doped {Mott}
  insulator}.
\newblock \emph{\bibinfo{journal}{Nature}} \textbf{\bibinfo{volume}{393}},
  \bibinfo{pages}{550--553} (\bibinfo{year}{1998}).

\bibitem{Hal00}
\bibinfo{author}{Halboth, C.~J.} \& \bibinfo{author}{Metzner, W.}
\newblock \bibinfo{title}{d-wave superconductivity and {Pomeranchuk}
  instability in the two-dimensional {Hubbard} model}.
\newblock \emph{\bibinfo{journal}{Phys. Rev. Lett.}}
  \textbf{\bibinfo{volume}{85}}, \bibinfo{pages}{5162--5165}
  (\bibinfo{year}{2000}).

\bibitem{Yam00}
\bibinfo{author}{Yamase, H.} \& \bibinfo{author}{Kohno, H.}
\newblock \bibinfo{title}{Possible quasi-one-dimensional fermi-surface in
  {La$_{2-x}$Sr$_x$CuO$_4$}}.
\newblock \emph{\bibinfo{journal}{J. Phys. Soc. Jpn.}}
  \textbf{\bibinfo{volume}{69}}, \bibinfo{pages}{332--335}
  (\bibinfo{year}{2000}).

\bibitem{Kee03}
\bibinfo{author}{Kee, H.-Y.}, \bibinfo{author}{Kim, E.~H.} \&
  \bibinfo{author}{Chung, C.-H.}
\newblock \bibinfo{title}{Signatures of an electronic nematic phase at the
  isotropic-nematic phase transition}.
\newblock \emph{\bibinfo{journal}{Phys. Rev. B}} \textbf{\bibinfo{volume}{68}},
  \bibinfo{pages}{245109} (\bibinfo{year}{2003}).

\bibitem{Ong03}
\bibinfo{author}{Wang, Y.} \emph{et~al.}
\newblock \bibinfo{title}{Dependence of upper critical field and pairing
  strength on doping in cuprates}.
\newblock \emph{\bibinfo{journal}{Science}} \textbf{\bibinfo{volume}{299}},
  \bibinfo{pages}{86--89} (\bibinfo{year}{2003}).

\bibitem{Fau06}
\bibinfo{author}{Fauque, B.} \emph{et~al.}
\newblock \bibinfo{title}{Magnetic order in the pseudogap phase of {high-$T_c$}
  superconductors}.
\newblock \emph{\bibinfo{journal}{Phys. Rev. Lett.}}
  \textbf{\bibinfo{volume}{96}}, \bibinfo{pages}{197001}
  (\bibinfo{year}{2006}).

\bibitem{Tra04}
\bibinfo{author}{Tranquada, J.~M.} \emph{et~al.}
\newblock \bibinfo{title}{Quantum magnetic excitations from stripes in copper
  oxide superconductors}.
\newblock \emph{\bibinfo{journal}{Nature}} \textbf{\bibinfo{volume}{429}},
  \bibinfo{pages}{534--538} (\bibinfo{year}{2004}).

\bibitem{Chr04}
\bibinfo{author}{Christensen, N.~B.} \emph{et~al.}
\newblock \bibinfo{title}{Dispersive excitations in the high-temperature
  superconductor {La$_{2-x}$Sr$_x$CuO$_4$}}.
\newblock \emph{\bibinfo{journal}{Phys. Rev. Lett.}}
  \textbf{\bibinfo{volume}{93}}, \bibinfo{pages}{147002}
  (\bibinfo{year}{2004}).

\bibitem{Hin04}
\bibinfo{author}{Hinkov, V.} \emph{et~al.}
\newblock \bibinfo{title}{Two-dimensional geometry of spin excitations in the
  high-transition-temperature superconductor {YBa$_2$Cu$_3$O$_{6+x}$}}.
\newblock \emph{\bibinfo{journal}{Nature}} \textbf{\bibinfo{volume}{430}},
  \bibinfo{pages}{650--653} (\bibinfo{year}{2004}).

\bibitem{Sch06}
\bibinfo{author}{Schnyder, A.}, \bibinfo{author}{Manske, D.},
  \bibinfo{author}{Mudry, C.} \& \bibinfo{author}{Sigrist, M.}
\newblock \bibinfo{title}{Theory for inelastic neutron scattering in
  orthorhombic {high-$T_c$} superconductors}.
\newblock \emph{\bibinfo{journal}{Phys. Rev. B}} \textbf{\bibinfo{volume}{73}},
  \bibinfo{pages}{224523} (\bibinfo{year}{2006}).

\bibitem{Sei05}
\bibinfo{author}{Seibold, G.} \& \bibinfo{author}{Lorenzana, J.}
\newblock \bibinfo{title}{Magnetic fluctuations of stripes in the high
  temperature cuprate superconductors}.
\newblock \emph{\bibinfo{journal}{Phys. Rev. Lett.}}
  \textbf{\bibinfo{volume}{94}}, \bibinfo{pages}{107006}
  (\bibinfo{year}{2005}).

\bibitem{Uhr04}
\bibinfo{author}{Uhrig, G.~S.}, \bibinfo{author}{Schmidt, K.~P.} \&
  \bibinfo{author}{Gr\"{u}ninger, M.}
\newblock \bibinfo{title}{Unifying magnons and triplons in stripe-ordered
  cuprate superconductors}.
\newblock \emph{\bibinfo{journal}{Phys. Rev. Lett.}}
  \textbf{\bibinfo{volume}{93}}, \bibinfo{pages}{267003}
  (\bibinfo{year}{2004}).

\bibitem{Voj06}
\bibinfo{author}{Vojta, M.}, \bibinfo{author}{Vojta, T.} \&
  \bibinfo{author}{Kaul, R.~K.}
\newblock \bibinfo{title}{Spin excitations in fluctuating stripe phases of
  doped cuprate superconductors}.
\newblock \emph{\bibinfo{journal}{Phys. Rev. Lett.}}
  \textbf{\bibinfo{volume}{97}}, \bibinfo{pages}{097001}
  (\bibinfo{year}{2006}).

\bibitem{Yao06}
\bibinfo{author}{Yao, D.~X.}, \bibinfo{author}{Carlson, E.~W.} \&
  \bibinfo{author}{Campbell, D.~K.}
\newblock \bibinfo{title}{Magnetic excitations of stripes and checkerboards in
  the cuprates}.
\newblock \emph{\bibinfo{journal}{Phys. Rev. B}} \textbf{\bibinfo{volume}{73}},
  \bibinfo{pages}{224525} (\bibinfo{year}{2006}).

\bibitem{Yam06}
\bibinfo{author}{Yamase, H.} \& \bibinfo{author}{Metzner, W.}
\newblock \bibinfo{title}{Magnetic excitations and their anisotropy in
  {YBa$_2$Cu$_3$O$_{6+x}$: Slave-boson} mean-field analysis of the bilayer
  {$t-J$} model}.
\newblock \emph{\bibinfo{journal}{Phys. Rev. B}} \textbf{\bibinfo{volume}{73}},
  \bibinfo{pages}{214517} (\bibinfo{year}{2006}).

\bibitem{Dai99}
\bibinfo{author}{Dai, P.} \emph{et~al.}
\newblock \bibinfo{title}{The magnetic excitation spectrum and thermodynamics
  of {high-$T_c$} superconductors}.
\newblock \emph{\bibinfo{journal}{Science}} \textbf{\bibinfo{volume}{284}},
  \bibinfo{pages}{1344--1347} (\bibinfo{year}{1999}).

\bibitem{Fon00}
\bibinfo{author}{Fong, H.~F.} \emph{et~al.}
\newblock \bibinfo{title}{Spin susceptibility in underdoped
  {YBa$_2$Cu$_3$O$_{6+x}$}}.
\newblock \emph{\bibinfo{journal}{Phys. Rev. B}} \textbf{\bibinfo{volume}{61}},
  \bibinfo{pages}{14773--14786} (\bibinfo{year}{2000}).

\bibitem{Sto04}
\bibinfo{author}{Stock, C.} \emph{et~al.}
\newblock \bibinfo{title}{Dynamic stripes and resonance in the superconducting
  and normal phases of {YBa$_2$Cu$_3$O$_{6.5}$ ortho-II}}.
\newblock \emph{\bibinfo{journal}{Phys. Rev. B}} \textbf{\bibinfo{volume}{69}},
  \bibinfo{pages}{014502} (\bibinfo{year}{2004}).

\bibitem{Bou00}
\bibinfo{author}{Bourges, P.} \emph{et~al.}
\newblock \bibinfo{title}{The spin excitation spectrum in superconducting
  {YBa$_2$Cu$_3$O$_{6.85}$}}.
\newblock \emph{\bibinfo{journal}{Science}} \textbf{\bibinfo{volume}{288}},
  \bibinfo{pages}{1234--1237} (\bibinfo{year}{2000}).

\bibitem{Esc05}
\bibinfo{author}{Eschrig, M.}
\newblock \bibinfo{title}{The effect of collective spin-1 excitations on
  electronic spectra in {high-$T_c$} superconductors}.
\newblock \emph{\bibinfo{journal}{Adv. Phys.}} \textbf{\bibinfo{volume}{55}},
  \bibinfo{pages}{47--183} (\bibinfo{year}{2006}).

\bibitem{Pai06}
\bibinfo{author}{Pailh\`{e}s, S.} \emph{et~al.}
\newblock \bibinfo{title}{Doping dependence of bilayer resonant spin
  excitations in {(Y,Ca)Ba$_2$Cu$_3$O$_{6+x}$}}.
\newblock \emph{\bibinfo{journal}{Phys. Rev. Lett.}}
  \textbf{\bibinfo{volume}{96}}, \bibinfo{pages}{257001}
  (\bibinfo{year}{2006}).

\bibitem{Zab06}
\bibinfo{author}{Zabolotnyy, V.~B.} \emph{et~al.}
\newblock \bibinfo{title}{Coexistence of metallicity and superconductivity in
  adjacent bilayers of a {high-$T_c$} superconductor}.
\newblock \emph{\bibinfo{journal}{arXiv:cond-mat}} \bibinfo{pages}{0608295}
  (\bibinfo{year}{2006}).

\bibitem{Tra95}
\bibinfo{author}{Tranquada, J.~M.}, \bibinfo{author}{Sternlieb, B.~J.},
  \bibinfo{author}{Axe, J.~D.}, \bibinfo{author}{Nakamura, Y.} \&
  \bibinfo{author}{Uchida, S.}
\newblock \bibinfo{title}{Evidence for stripe correlations of spins and holes
  in copper oxide superconductors}.
\newblock \emph{\bibinfo{journal}{Nature}} \textbf{\bibinfo{volume}{375}},
  \bibinfo{pages}{561--563} (\bibinfo{year}{1995}).

\bibitem{Kiv03}
\bibinfo{author}{Kivelson, S.~A.} \emph{et~al.}
\newblock \bibinfo{title}{How to detect fluctuating stripes in the
  high-temperature superconductors}.
\newblock \emph{\bibinfo{journal}{Rev. Mod. Phys.}}
  \textbf{\bibinfo{volume}{75}}, \bibinfo{pages}{1201--1241}
  (\bibinfo{year}{2003}).

\bibitem{And02}
\bibinfo{author}{Ando, Y.}, \bibinfo{author}{Segawa, K.},
  \bibinfo{author}{Komiya, S.} \& \bibinfo{author}{Lavrov, A.~N.}
\newblock \bibinfo{title}{Electrical resistivity anisotropy from self-organized
  one dimensionality in high-temperature superconductors}.
\newblock \emph{\bibinfo{journal}{Phys. Rev. Lett.}}
  \textbf{\bibinfo{volume}{88}}, \bibinfo{pages}{137005}
  (\bibinfo{year}{2002}).

\bibitem{And05}
\bibinfo{author}{Andersen, B.~M.} \& \bibinfo{author}{Hedeg{\aa}ard, P.}
\newblock \bibinfo{title}{Spin dynamics in the stripe phase of the cuprate
  superconductors}.
\newblock \emph{\bibinfo{journal}{Phys. Rev. Lett.}}
  \textbf{\bibinfo{volume}{95}}, \bibinfo{pages}{037002}
  (\bibinfo{year}{2005}).

\end{thebibliography}

\section*{Supplementary material}

\begin{figure}[h!]
\includegraphics[width=9cm]{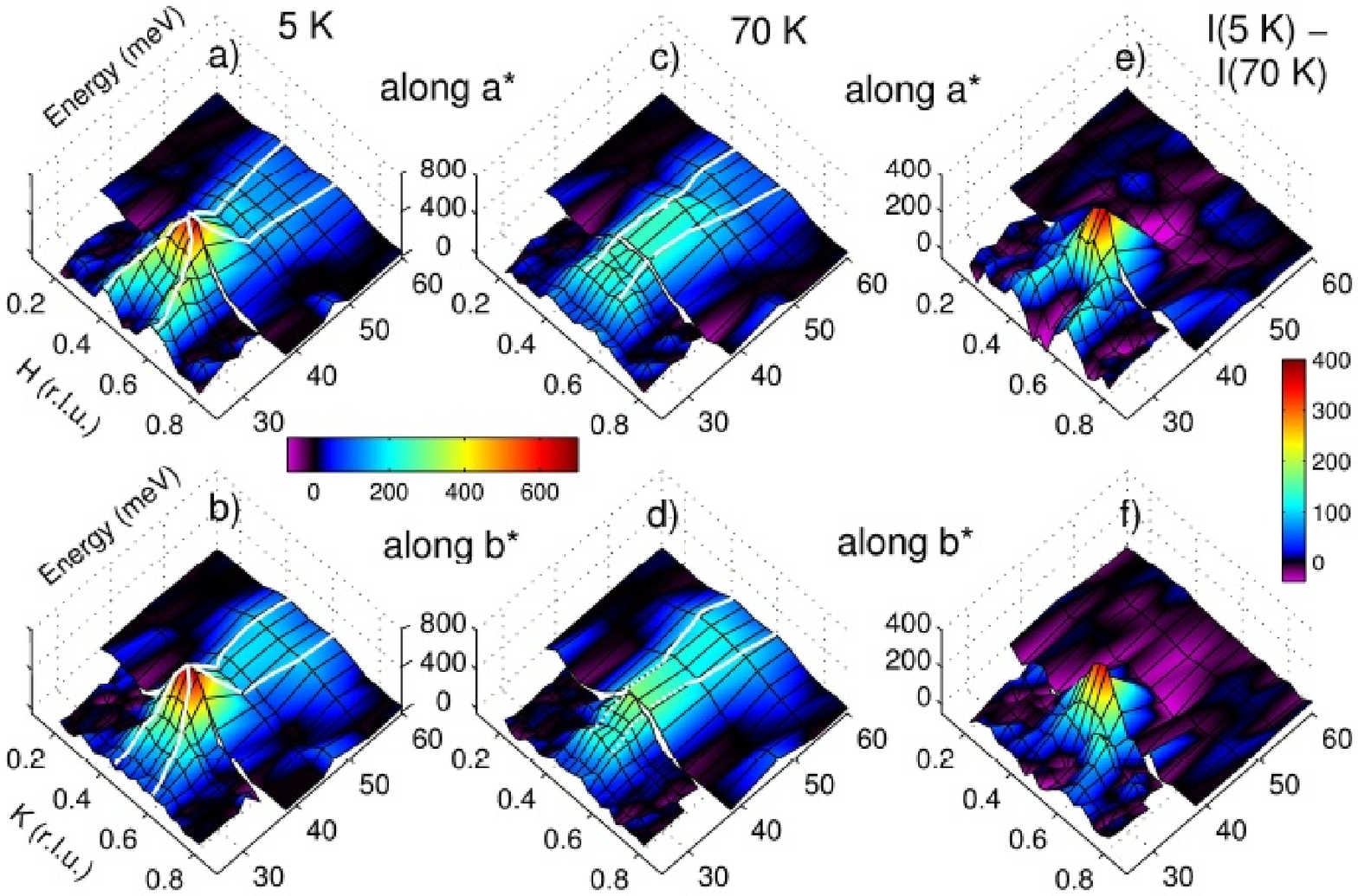}
\\[10pt]
Figure S1: Colour representation of the magnetic intensity. Panels \textbf{a,b} show the SC regime, \textbf{c,d} the regime just above $T_c$, and \textbf{e,f} the difference between both spectra. The upper and lower rows show scans along the \astar-axis ($H$, -1.5, -1.7) and \bstar-axis (1.5, $K$, 1.7),
respectively. In order to obtain a meaningful colour representation, the
intensity at 250 K was subtracted for $E <$ 38~meV and the data was corrected for a \textbf{Q}-linear background at all energies. The final wave vector was fixed to $2.66\, \mbox{\AA}^{-1}$ below 38~meV and to $4.5\, \mbox{\AA}^{-1}$ above. In contrast to Fig. 2 of the manuscript, where the colour scale was normalized to the peak intensity of the scan at each individual energy, we use an energy-independent colour scale here. Scans taken at the overlapping energy 38~meV were used to normalize both data sets. Crossings of black lines represent measured data points. White lines connect the fitted peak positions of the constant-energy cuts. Dotted lines represent upper bounds on the incommensurability.

\label{figS1}
\end{figure}

In Fig. S1, the data in the superconducting (5~K) and pseudogap (70~K) states (same as in Fig. 2 of the manuscript) are reproduced in a different presentation scheme with a global colour normalization. The difference spectrum $I(5\text{
K})-I(70\text{ K})$ is also shown. The difference signal comprises the downward-dispersing branch below the resonance energy $\Eres = 37.5\text{ meV}$, which draws its spectral weight from a limited range above and a
more extended range below \Eres\ (negative signal in Fig. S1 \textbf{e},\textbf{f}). It
is significantly less anisotropic than the PG spectrum itself (Fig.
S1 \textbf{c},\textbf{d}), with respect to both the incommensurability  \IC\ and the spectral weight
distribution. Notably, the difference spectrum is very similar to its
analog in almost optimally doped $\rm YBa_2 Cu_3 O_{6.85}$, which was
shown to exhibit a nearly circular geometry in V. Hinkov et al.,
$Nature$ \textbf{430}, 650.

Fig. S2 highlights again the difference between the ``hour glass" and
``vertical" dispersion relations in the SC and PG states,
respectively. In the SC state, the profile along \astar\ exhibits a
single peak at $\Eres = 37.5$ meV, corresponding to the neck of the
``hour glass". In the PG state, the profile at this energy consists
of two incommensurate peaks. The opposite trend holds along \bstar\
at energies below \Eres. The profile along \bstar\ is very narrow in
the PG state, and incommensurate reflections cannot be resolved
(dashed lines in Fig. 3 of the manuscript). Below \Eres, the
incommensurability increases in the SC state, where the anisotropy
between the dispersions along \astar\ and \bstar\ is significantly
reduced.

\begin{figure}[b]
\includegraphics[width=8cm]{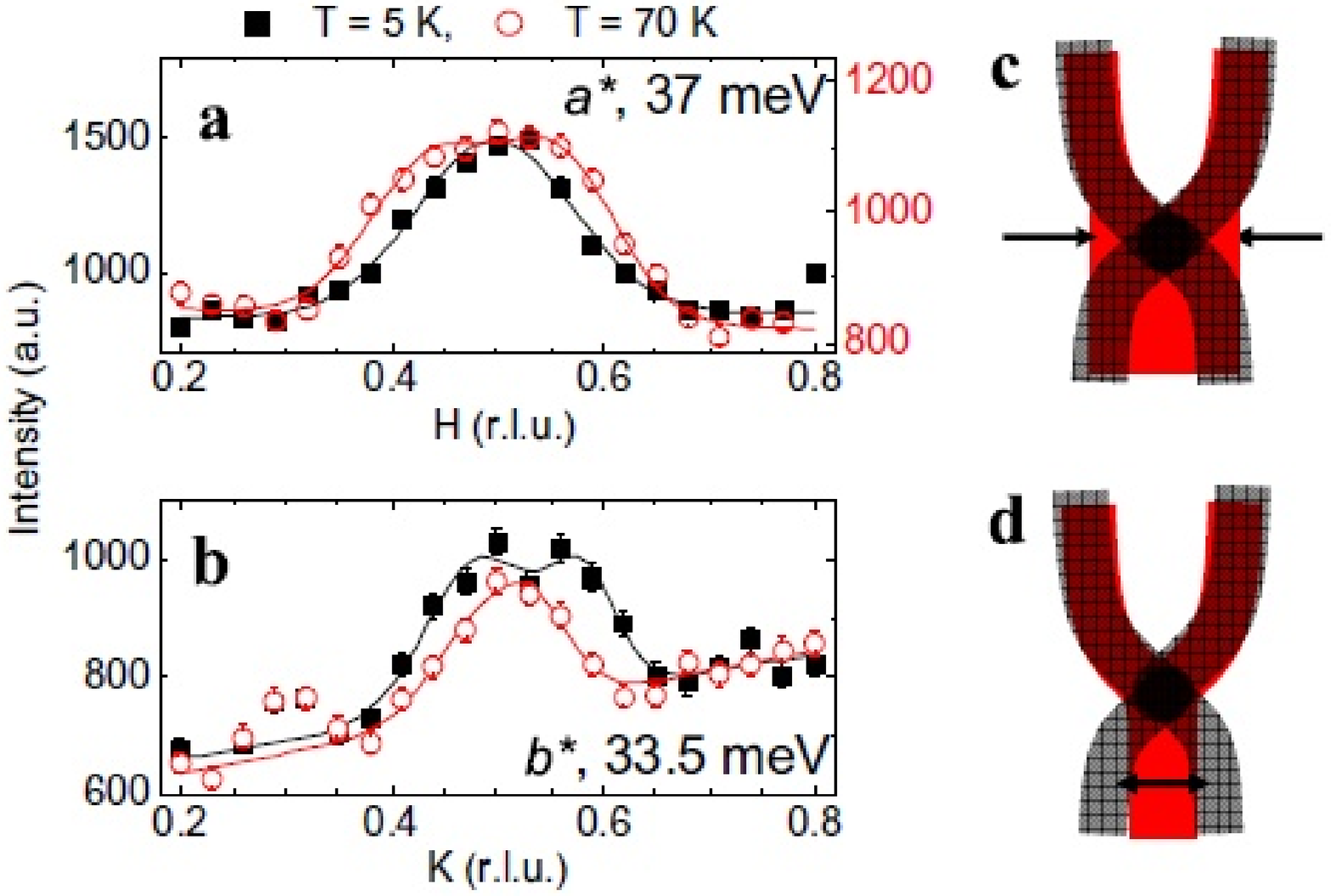}
Figure S2: Comparison of constant-energy profiles at 37 meV along \astar, panel \textbf{a}, and at 33.5 meV along \bstar, panel \textbf{b}, in the SC and PG states. The data was taken from Figs. 1\textbf{e} and \textbf{h} of the manuscript. At 37~meV, the amplitudes of both scans were scaled to the same amplitude in order to compare their $Q$-width. Panels \textbf{c} and \textbf{d} show corresponding sketches of the dispersion relations in the SC state (grey) and the PG state (red).
\label{figS2}
\end{figure}

Fig. S3 shows low-energy profiles deep in the SC state, along with
data immediately below and above the SC transition temperature. While
the scans at 5 and 50 K show clear incommensurate peaks are identical
within the experimental error, the profile in the scan at 70 K is
flat-top, and the peak intensity is considerably reduced. The data
are complementary to the temperature sweeps at constant wave vector
presented in Fig. 4 of the manuscript and demonstrate the abrupt
modification of the magnetic spectrum at the SC transition
temperature.

\begin{figure}[]
\includegraphics[width=6cm]{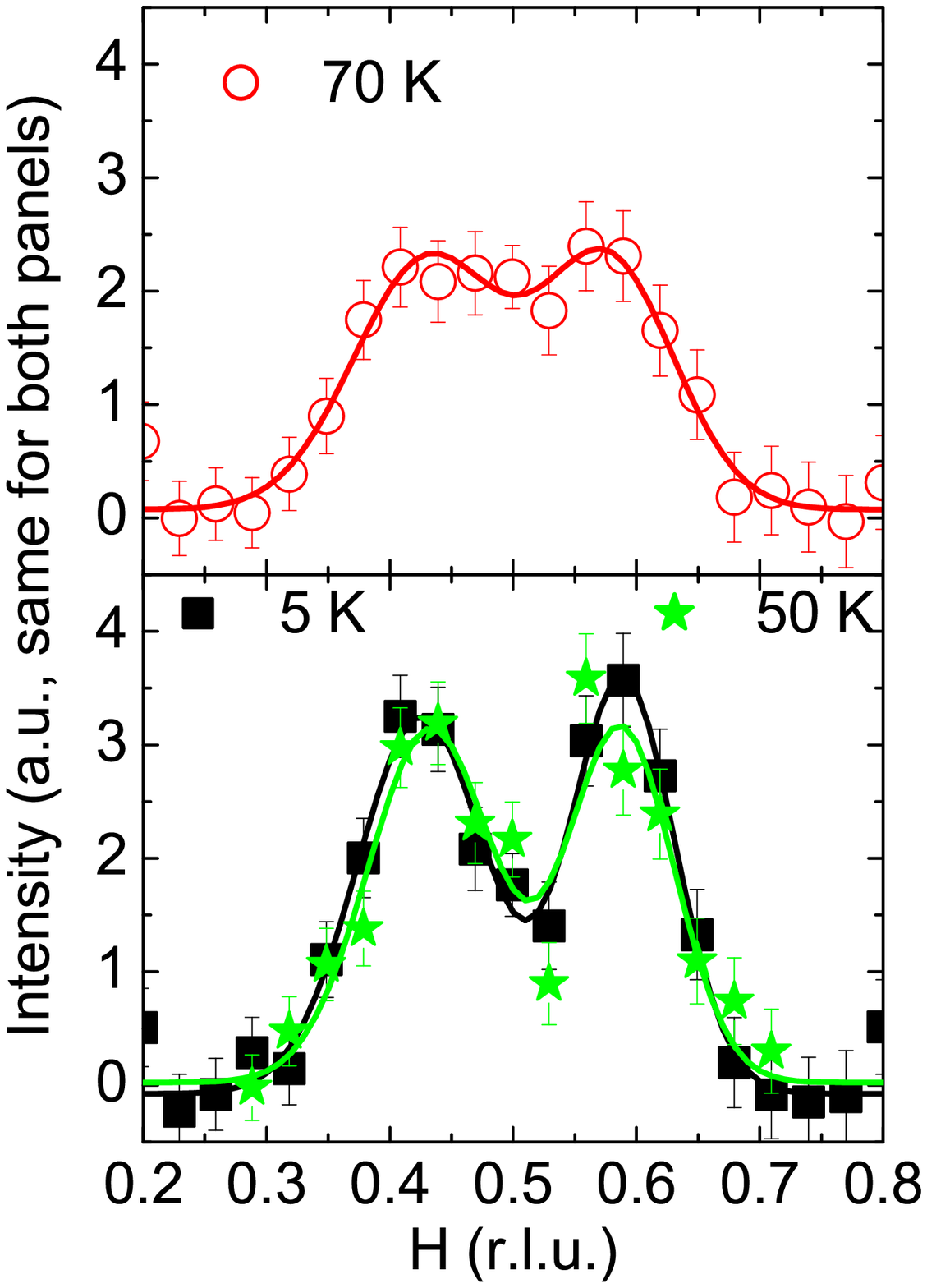}

Figure S3: Constant-$E$ scans at 31~meV along \astar\ at different temperatures. For better comparison, a background measured at room temperature has been subtracted.
\label{figS3}
\end{figure}

Fig. S4 provides a synopsis of the spectra in the SC state at 5 K,
the PG state at 70 K, and the normal state at room temperature,
integrated along the wave vector component along \astar. Whereas all
three spectra are similar at the highest energies, they exhibit
qualitative differences at lower energies, as described in the manuscript.

\begin{figure}[b]
\includegraphics[width=7cm]{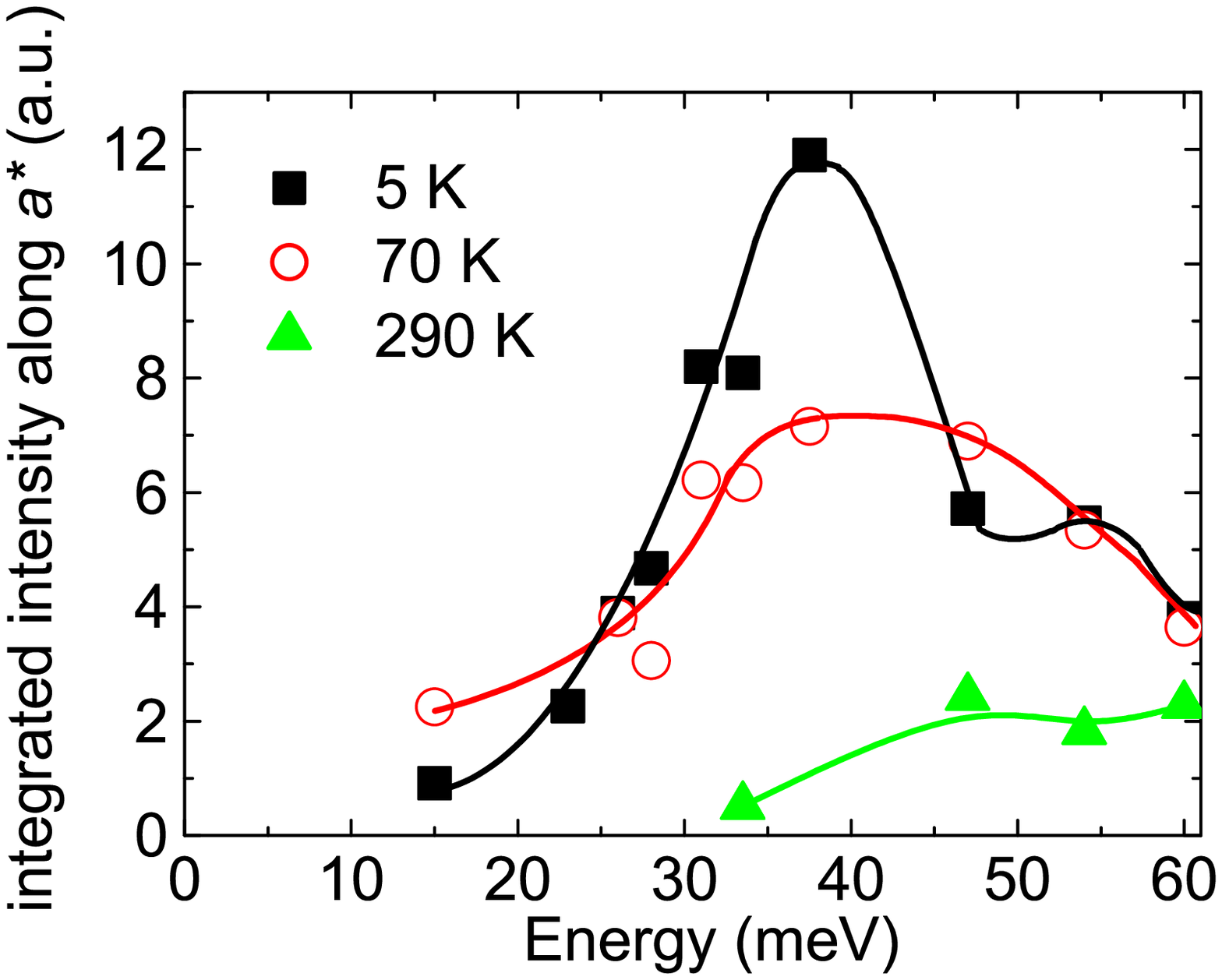}

Figure S4: Energy dependence of the integrated intensity along \astar\ in the SC state (5~K), the PG state (70~K) and the normal state (290~K).
\end{figure}

\end{document}